\documentclass[doublespacing]{elsart}
\usepackage{amssymb}
\usepackage{amsfonts}
\usepackage{graphicx}

\begin{document}

\begin{frontmatter}



\title{Surrogate data method applied to nonlinear time series}


\author{Xiaodong Luo,}\ead{enxdluo@eie.polyu.edu.hk}
\author{Tomomichi Nakamura,}\ead{entomo@eie.polyu.edu.hk}
\and
\author{Michael Small}\ead{ensmall@polyu.edu.hk}

\address{Department of Electronic and Information Engineering, The Hong Kong Polytechnic
University, Hung Hom, Hong Kong.}

\begin{abstract}
The surrogate data method is widely applied as a data dependent
technique to test observed time series against a barrage of
hypotheses. However, often the hypotheses one is able to address
are not those of greatest interest, particularly for system known
to be nonlinear. In the review we focus on techniques which
overcome this shortcoming. We summarize a number of recently
developed surrogate data methods. While our review of surrogate
methods is not exhaustive, we do focus on methods which may be
applied to experimental, and potentially nonlinear, data. In each
case, the hypothesis being tested is one of the interests to the
experimental scientist.
\end{abstract}

\end{frontmatter}

\section{Introduction}

\subsection{Overview}

\begin{figure}[!t]

\centering

\includegraphics[width=4.5in]{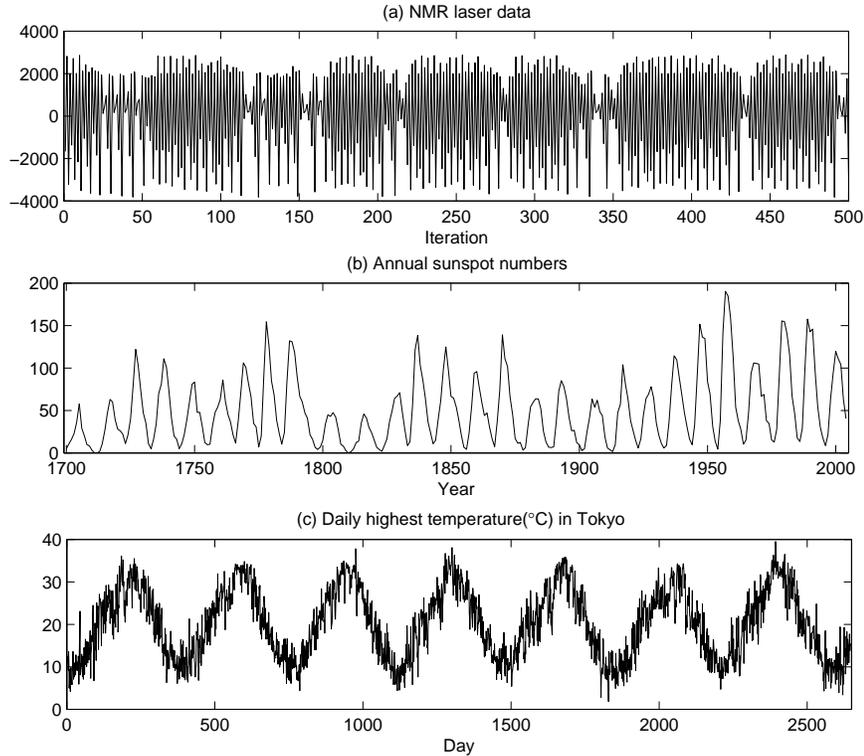}

\caption{(a) Nuclear magnetic resonance~(NMR) laser data, (b)
Annual sunspot numbers from the year 1700 to 2004, and (c) Daily
highest temperature in Tokyo from 1 January 1998 to 31 March
2005.}

\label{somedata}

\end{figure}

The physical phenomena in the real world are usually attributed to
certain causalities \cite{gousebook}. With different causalities,
the corresponding phenomena, often captured in the time series of
measurement, might diverge significantly as illustrated in Fig.
\ref{somedata}: They could be irregular fluctuations as shown in
panel (a), or pseudo-periodic data in panel (b), or the mixture,
i.e., irregular fluctuations with periodic trends as indicated in
panel (c).

To understand the underlying mechanisms responsible for generating
those different time series in Fig. \ref{somedata}, one needs to
reply some elementary questions first: are the data linear or
nonlinear, stochastic or deterministic, pseudo-periodic or
chaotic? To explore the possible answers, the method of Monte
Carlo hypothesis test \cite{Barnard discussion,Hope simplified},
or surrogate data test equivalently \cite{Galka topics,Schreiber
surrogate,Small applied,Theiler testing,Theiler constrained}, is
often applied. This method has become a useful tool to validate
the results of dynamical analysis, and thus help understand the
causal processes underlying the experimental data. For example, if
through the method one finds that irregular fluctuations are not
random variables, then it immediately implies that there exists
some kind of dynamical (deterministic) structure. Therefore a
consequential conclusion is that, it is possible to build
deterministic models~(or model systems) from the time series.

The focus of this section is to introduce the framework of Monte
Carlo hypothesis test. Moreover, we will discuss some important
concepts associated with the components that form the framework,
which will be frequently applied in our later analysis.

\subsection{Monte Carlo hypothesis test}

Null hypothesis tests use statistical measures of the underlying
system to determine the probability that a proposed hypothesis is
true (or false) \cite{Weisstein hypothesis}. The common procedures
include \cite{Good permutation}:

\begin{enumerate}
\item Formulate the null hypothesis of interest, the alternative hypothesis
and the potential risks associated with a decision.

\item Choose a test statistic.

\item Compute the frequency distribution of the test statistic under the
null.

\item With the guide of the frequency distribution, choose certain criterion
to determine whether to reject the hypothesis or not.
\end{enumerate}

It is easy to see that the framework of hypothesis testing
consists of the null hypothesis, the test statistic, the frequency
distribution of the test statistic, and the discriminating
criterion that determines whether to reject the null hypothesis or
not. In order to obtain the frequency distribution of the test
statistic, traditionally one would need to carefully choose the
test statistic such that it follows a well known distribution. But
in practice, on one side it might be difficult to find the refined
statistics for tests in many situations; on the other, with modern
computers the computer-intensive methods become feasible and
popular \cite{Noreen computer}. Hence, the concept of Monte Carlo
hypothesis test naturally appeared \cite{Barnard discussion,Hope
simplified}. The basic idea is to produce a number of different
realizations under the null through Monte Carlo simulation. In
practice these realizations are usually generated from the
original experimental data, but not really ever observed,
therefore they will also be called the surrogate data in the
later. From the ensemble of the surrogates, one could calculate
the empirical distribution and the confidence interval of the test
statistic. In this sense, the frequency distribution will
essentially depend on the surrogate generation algorithm (and the
chosen statistic of course). Therefore one could also say that the
surrogate algorithm is one of the elements that form a null
hypothesis test, as we do in the later.

\subsection{Pivotality and constrained-realization surrogates}

For the convenience of our later discussion, we need to introduce some
terminologies. Following the notation in \cite{Small applied,Theiler
constrained}, let $\mathcal{F}$ be the set of all the possible processes for
the problem under consideration. Also let $\phi$ be the formulated null
hypothesis and $\mathcal{F}_{\phi}$ the set of processes that are consistent
with the null $\phi$. If $\mathcal{F}_{\phi}$ consists of only one element,
then the null hypothesis $\phi$ is called {\em simple}, otherwise it is called
{\em composite}.

Given a composite null hypothesis $\phi$ and a process $F$ consistent with $%
\phi$, let us denote the chosen test statistic by $T$, and the corresponding
probability distribution function (PDF) under the null hypothesis by $%
P_{T,F}(t)\equiv Prob(T<t|F \in \mathcal{F}_{\phi})$. If for any two
processes $F_i$ and $F_j$ ($i \neq j$) in the set $\mathcal{F}_{\phi}$, one
has that $P_{T,F_i}(t)=P_{T,F_j}(t)$, then the statistic $T$ is said
{\em pivotal}; otherwise it is {\em non-pivotal}.

A remarkable advantage of test statistics which are pivotal, as
can be seen from the definition, is that one will always obtain
the same statistic distribution $P_{T}(t)$, which is independent
of the process $F$ chosen from the set $\mathcal{F}_{\phi}$.
Therefore adopting a pivotal test statistic might significantly
reduce the difficulty in devising the algorithm to produce
surrogates of the null \cite{Small correlation}. However, if the
test statistic is non-pivotal, then there is no guarantee that
$P_{T,F_i}(t)=P_{T,F_j}$ holds for arbitrary processes $F_i$ and
$F_j$ in $\mathcal{F}_{\phi}$. Suppose that the time series
$\mathbf{d}=\{d_i\}_{i=1}^{n}$ under test is generated from a
process $F_d \in \mathcal{F}_{\phi}$, then in order to avoid
possible false
rejection of the null hypothesis \footnote{%
For example, there is a process $F_r \in \mathcal{F}_{\phi}$, but $%
P_{T,F_r}(t) \neq P_{T,F_d}(t)$. With $P_{T,F_d}(t)$ as the reference
distribution, one would falsely reject the hypothesis.}, it is usually
required, as the sufficient condition, that the process $F_s$ producing the
surrogate data satisfies $F_s=F_d$ \footnote{%
Or more loosely, $\hat{F}_s=\hat{F}_d$ \cite{Small applied}. Here
$\hat{F}_s$ denotes the process estimated from the surrogate and
$\hat{F}_d$ the process estimated from the original data.
Strictly, the condition that the estimates coincide rather than
the processes themselves helps to prevent to possibility over
over-constrained surrogates. Such over-constrainedness generates
surrogates that agree to closely with the true data, and the
discriminating power of the hypothesis test is reduced \cite{Small
applied}.}. Surrogates generated from such processes are called
{\em constrained} realizations; otherwise it is said {\em
non-constrained}. Obviously, given a set of
constrained-realization surrogates $\{\mathbf{s_1, s_2, ...}\}$,
any adopted test statistic $T$ will appear as if it were pivotal
for the processes $\{F_d, F_{s_1}, F_{s_2},... \}$.

In the following sections, we will introduce various applications
of the Monte Carlo hypothesis test method. One of the popular
applications is to detect nonlinearity in a time series, as
described in \cite{Nakamura testing,Takens detecting,Theiler
testing,Theiler constrained}. Other applications include the
detection of aperiodicity \cite{Luo surrogate,Small
surrogate,Theiler on} and the correlation between irregular
fluctuations with long term trends \cite{Nakamura shuffle}. As an
important, and possibly the most attractive component, the
well-tailored surrogate generation algorithm associated with the
hypothesis deserves to receive great attention. In fact, because
of its importance, the Monte Carlo hypothesis test method is often
called surrogate data test, or surrogate data method \cite{Galka
topics,Schreiber surrogate,Small applied} in the literature. In
this review we use these two terms interchangeably.

There are already some excellent introductory works covering the
topic of the surrogate data method (for example \cite{Galka
topics,Schreiber surrogate,Small applied}), therefore we will not
provide excessive detail in this paper. Instead, we will dedicate
our effort to introducing some of more recent progress. The
readers are referred to the broad literature, much of which is
cited here, for further detail. The three reviews \cite{Galka
topics,Schreiber surrogate,Small applied} are particularly
recommended.

\section{Surrogate test for detection of nonlinearity}

A rational step before the application of nonlinear time series
methods is to identify the presence of nonlinearity. For this
purpose, one could employ the direct detection strategy, that is,
in order to detect nonlinearity, one adopts some characteristic
nonlinear statistics, such as the correlation dimension, the
Lyapunov exponent, the continuity, and so on \cite{Kaplan
direct,Kaplan exceptional,Wayland}, as the discriminating measures
in the belief that these statistics reveal the essential behaviors
of nonlinear systems. This strategy, however, may encounter a few
disadvantages in practice. On one hand, in certain situations the
characteristic nonlinear statistics do not play the role well as
the unequivocal identifiers of the underlying systems \cite{Small
applied}. Take the correlation dimension \cite{Grassberger
characterization,Grassberber measure} as an example, it was shown
that some linear stochastic processes with simple power law
spectra would also have finite non-integer values as many
nonlinear systems do \cite{Osborne finite}, thus one would fail to
distinguish between linearity and nonlinearity by simply examining
the values of the correlation dimension. On the other hand, given
only a limited amount of the realizations of the underlying
systems, it is often difficult to evaluate the reliability of the
test results based on the direct detection strategy. The situation
becomes even worsen with the presence of noise, which, often is
the case, will reduce the discriminating power of the
characteristic statistics. Thus one is forced back to the
aforementioned scenario, i.e., the adopted characteristic
statistic fails to unequivocally identify the underlying system.
As an example, let us consider the (largest) Lyapunov exponent.
Theoretically its value shall be zero for a periodic orbit.
However, if the periodic orbit is perturbed by noise components,
the value of the Lyapunov exponent might slightly increase and
become positive, which is often deemed as the sign of chaos and
therefore possibly engenders misleading conclusion.

An alternative strategy for nonlinearity detection is the surrogate data
methods, as an application of the Monte Carlo hypothesis test. The test
procedures go as follows, one first proposes a null hypothesis which usually
assumes that the time series is initially generated by a linear stochastic
process. With the null hypothesis, one produces an ensemble of surrogate
data based on the original time series. Then one chooses a proper test
statistic in the sense that, if the original data is consistent with the
null hypothesis, the statistic of the original data shall follow the same
distribution as those of the surrogates, otherwise it shall appear atypical
to the distribution. After calculating the test statistic, one inspects
whether the statistic value of the original data appears typical to the
distribution of the surrogates according to certain discriminating
criterion. If the answer is no, one rejects the null with certain confidence
level (depending on the chosen discriminating criterion, as will be
discussed in the later), which implies that the data in test is very likely
to be nonlinear.

In this section, we will first review the the hierarchical surrogate data
tests for nonlinearity detection proposed by Theiler et al. \cite{Theiler
testing,Theiler constrained}. We will also introduce other surrogate data
methods \cite{Nakamura testing,Schreiber improved,Schreiber constrained},
which essentially follow the same hierarchical framework but differentiate
in the way of surrogate generation, as will be explained in the following.

\subsection{The hierarchical surrogate data tests}

Here we confine our discussion to stationary irregular time series. With the
data, as the first step of analysis we want to detect the potential
nonlinearity in it. Of course it is also possible that the irregular data is
produced from a linear stochastic system, or a linear deterministic system
contaminated by noise. Since the stationary data generated by a linear
deterministic system appear either periodic or constant, even with noise
components it is trivial to distinguish between linear stochastic and
deterministic cases if the noise level is not extremely high \footnote{%
A constant with (either linear or nonlinear) stochastic noise can be
considered as a stochastic case; while a periodic orbit perturbed by noise
will still have long-term linear correlations, which are usually not
possessed by stationary linear stochastic processes.}. Therefore in the
later we will only consider the scenario that the stationary irregular time
series is generated from either a linear stochastic process or a nonlinear
stochastic or deterministic process. Following the aforementioned
procedures, we will introduce one by one the basic elements that form the
framework of null hypothesis tests.

\subsubsection{Null hypotheses}

The basic assumption is that the time series under test is from a
linear stochastic noise process, either i.i.d. (independent and
identically distributed) or correlated (in the general form of an
auto-regressive moving average $ARMA$ process). But note that, it
is possible to introduce nonlinearity into the original linear
data during the measurement step by letting the original data pass
through a nonlinear filter \cite{Schreiber surrogate}. With this
consideration, one could formulate the following hierarchical
composite hypotheses (or those equivalently stated in \cite{Galka
topics,Small applied,Theiler testing,Theiler constrained}), as
shown in Fig. \ref{HierarchicalNH}:

\begin{figure}[!t]

\centering

\includegraphics[width=3in]{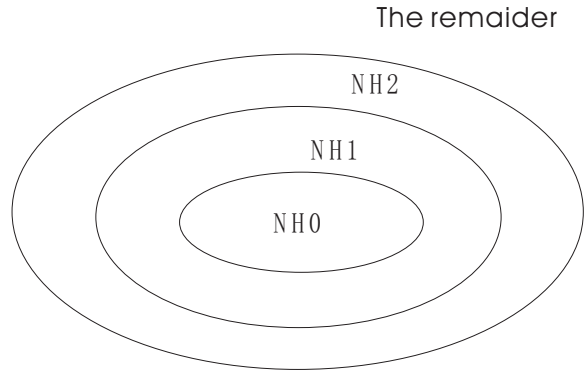}

\caption{Schematic representation of the hierarchical hypotheses
and their alternatives.}

\label{HierarchicalNH}

\end{figure}

\begin{itemize}

\item {\bf Null Hypothesis 0 (NH0)}: The data in test are i.i.d.
noise with unknown mean and variance.

\item {\bf Null Hypothesis 1 (NH1)}: The data in test are produced
from a linear stochastic process in the form of an $ARMA$ model
with unknown parameters, which is essentially a linear filter of
the i.i.d noise.

\item {\bf Null Hypothesis 2 (NH2)}: The data in test are obtained
by applying a static and monotonic nonlinear filter to the time
series originally generated by an $ARMA$ process.

\end{itemize}

\subsubsection{Test statistic}

In principle one shall choose the statistics in a way that
measures of the original data and the surrogates shall appear
consistent if the null hypothesis is true; otherwise they will
reveal the discrepancy. For this purpose, one of the popular
choices in the literature is the correlation dimension, since it
was shown that the correlation dimension is a pivotal statistic
for the hierarchical null hypothesis tests for nonlinearity
detection \cite{Small correlation,Small applied}. Of course, there
are also many other proper candidates. And in general, the
discriminating powers of the test statistics may vary from case to
case \cite{Schreiber discrimination}.

\subsubsection{Surrogate generation algorithms}

Broadly speaking, there are two strategies to generate surrogate
data. One is to first build a parametric model based on the null
hypothesis (rather than the original data), and then use the model
to produce the surrogates. The other one is to seek a
nonparametric model to produce surrogates consistent with the null
hypothesis, which is especially useful for the test of a composite
null hypothesis and will thus be the focus in our later
discussions. Of course, parametric algorithms can also be
constructed for composite hypotheses as long as one could find
suitable pivotal test statistics. In such situations, the
hypothesis which one can test is that, the data are consistent
with a particular parametric model (possibly fitted to the data),
or any other model with the same frequency distribution of the
test statistic. This parametric approach can be particularly
useful if one is interested in providing a behavior (or dynamics)
based test of the suitability of a particular model. However, such
parametric algorithms are often non-constrained and we will not
consider them in this review. The interested readers are referred
to \cite{Galka topics,Small correlation, Takens detecting, Small
correlation}.

Now let us begin introducing the nonparametric (and constrained)
surrogate generation algorithms that correspond to the above
hierarchical null hypotheses:

\begin{itemize}

\item {\bf Algorithm 0}: To produce i.i.d. surrogates, one only
needs to randomly shuffle the original data.

\item{\bf Algorithm 1}: To produce linear stochastic surrogates,
one first applies the Fourier transform to the original data to
obtain the corresponding moduli and phases. Then one keeps the
moduli but replace the phases by random numbers uniformly drawn
from the interval $(-\pi,\pi]$. Finally one applies the inverse
Fourier transform to the coefficients with preserved moduli but
randomized phases. Thus obtained data are the desired surrogates.

\item{\bf Algorithm 2}: To produce surrogates consistent with {\it
NH2}, one first needs to invert the static and monotonic nonlinear
filter \footnote{The surrogate generation algorithm could also be
extended to the cases with non-monotonic filters, as pointed out
by Kuiumtzis \cite{Kugiumtzis surrogate}} to obtain the original
linear stochastic data, then applies {\it Algorithm 1} to generate
interim surrogate data of the linear data, and finally introduce
the nonlinear filter back into the interim surrogates to obtain
the final surrogates.

\end{itemize}

The surrogates produced from the above algorithms will preserve
the amplitude distribution of the original data as we expect.
However there also exist a few defects in practice. One problem is
that, for surrogates generated by {\it Algorithm 1}, their power
spectra will often deviate from that of the original data. As a
remedy, Schreiber and Schmitz \cite{Schreiber improved} suggested
to repeat the surrogate generation procedures until the difference
between the power spectra reaches certain stopping criterion.
Another problem is that, to apply the discrete Fourier transform,
the data has to be assumed periodic. Therefore the wraparound
artifact \cite{Galka topics,Theiler reexamination} will be
introduced. A possible remedy, as suggested in \cite[pp.
238-240]{Galka topics}, is to conduct limited phase randomization.
Another remedy is to avoid adopting the Fourier transform, as will
be discussed later.

\subsubsection{Discriminating criterion}

Since the exact knowledge of the statistic distribution is often
not available, one will resort to certain discriminating criterion
to help make the decision and determine the corresponding
confidence level (if to reject). The popular discriminating
criteria in the literature include two classes: parametric and
nonparametric. The parametric criterion assumes that the statistic
follows a Gaussian distribution, and the distribution parameters,
i.e. the mean and the variance, would be estimated from the finite
samples. One can determine whether to reject the null by examining
whether the statistic of the original time series follows the
statistic distribution of the surrogates, while the corresponding
confidence level of inference can be calculated from the estimated
statistic distribution; The nonparametric criterion \cite{Theiler
using} examines the ranks of the statistic values of the original
time series and its surrogates. Supposes that the statistic of the
original time series is $T _{0}$ and the surrogate values are $\{T
_{i}\}_{i=1}^{N}$ given $N$ surrogate realizations. Then if the
statistic of both the original time series and the surrogates
follows the same distribution, the probability is $ 1/(N+1)$ for
$T _{0}$ to be the smallest or largest among all of the values
$\{T _{0},T _{1},...,T _{N}\}$. Thus if $N$ is large, when one
finds that $T _{0}$ is smaller or larger than all of the values in
$\{T _{i}\}_{i=1}^{N}$, it is quite possible that $T _{0}$ instead
follows a different distribution from that of $\{T
_{i}\}_{i=1}^{N}$. Hence the criterion rejects the null hypothesis
whenever the original statistic $T _{0}$ is the smallest or
largest among $\{T _{0},T _{1},...,T _{N}\}$, the false rejection
rate is considered as $1/(N+1)$ for one-sided tests and $2/(N+1)$
for two-sided ones.

\subsection{Other methods to generate surrogates}

\subsubsection{The temporal shift algorithm}

The basic idea of the temporal shift algorithm \cite{Luo
surrogate,Nakamura testing} goes as follows: For two independent
time series ${\bf x}=\{x_i\}_{i=1}^N$ and ${\bf
y}=\{y_i\}_{i=1}^n$, if they are produced from a same linear
stochastic process, then the additions ${\bf z}=\{z_i: z_i=\alpha
x_i+\beta y_i\}_{i=1}^N$ for arbitrary real scalar coefficients
$\alpha$ and $\beta$ will also follow the same linear process,
although possibly with different initial conditions. However, if
${\bf x}$ and ${\bf y}$ are from a nonlinear (stochastic or
deterministic) process, in general adding them together will
increase the complexity. Thus their additions ${\bf z}$ may behave
different from ${\bf x}$ and ${\bf y}$. And by adopting a proper
test statistic, one may detect this difference.

In practice, if only given a single time series ${\bf
d}=\{d_i\}_{i=1}^N$, then in order to produce surrogates, one
could extract two subsets from the original data, for example,
${\bf d_1}=\{d_i\}_{i=1}^{N-\tau}$ and ${\bf
d_2}=\{d_i\}_{i=1+\tau}^{N}$, where parameter $\tau$ is the
temporal shift-or more precisely, index shift-between ${\bf d_1}$
and ${\bf d_2}$, and is often required to decorrelate ${\bf d_1}$
and ${\bf d_2}$. The surrogates ${\bf s}$ are produced according
to the formula $s_i=\alpha d_i+\beta d_{i+\tau}
(i=1,2,...,N-\tau)$, by either varying the temporal shift $\tau$,
or randomizing the coefficients $\alpha$ and $\beta$, or the
combinations. In principle, there is no requirement for the
coefficients $\alpha$ and $\beta$. But note that, if the ratio
$\alpha/\beta \rightarrow 0$ or $\infty$, then roughly the
surrogates ${\bf s}=\alpha {\bf d_1}+\beta {\bf d_2} \propto {\bf
d_2}$ or ${\bf d_1}$. Therefore whether the original data is
consistent the null hypothesis or not, the produced surrogates
will look very close to it. Consequently, even if the null
hypothesis does not actually hold, the test statistic may fail to
detect the tiny difference between the surrogates and the original
data. Thus spurious results will appear in these situations. To
avoid this problem, we suggest that the ratio $\alpha/\beta$ takes
moderate values. For detail, see \cite{Luo surrogate}.

From the above discussions, it is easy to find that the temporal
shift algorithm actually utilizes the fact that the superposition
principle is applicable to linear processes rather than nonlinear
ones. And with this fact, one could avoid applying the Fourier
transform to the original data and thus circumvent the
consequential wraparound artifact. One may also note that, in
contrast to the standard algorithms (i.e., {\it Algorithm 0-2} and
the iterative version of {\it Algorithm 2}) \footnote{The temporal
shift algorithm only produces surrogates for {\it NH0-1}, but one
could naturally extend it to producing surrogates for {\it NH2}.
For this purpose, we suggest that one only inverts the nonlinear
filter and then compares the interim surrogates to the inverted
original data. It might cause problems to introduce the nonlinear
filter back into the interim surrogates in the same way of {\it
Algorithm 2}, since the temporal shift algorithm does not exactly
preserve the amplitude distribution of the original data.}, the
surrogates produced by the temporal shift algorithm do not exactly
preserve the amplitude distribution of the original data. However,
since the algorithm ensures that the surrogates are generated from
the same process under the null hypothesis, it is still a
constrained-realization surrogate generation algorithm. In our
viewpoint, the elimination of the restriction to preserving the
amplitude distribution is actually an advantage of the new
algorithm, which makes the algorithm more flexible and efficient
to produce surrogates.

\subsubsection{The simulated annealing method}

Simulated annealing \cite{Kirkpatrick optimization} is a
stochastic approach that mimics the physical process to solve the
combinatorial optimization problems. Physically, the annealing
process starts from a high temperature that melts the solids, then
one gradually decreases the temperature. If the variation
amplitude of the temperature is small enough, after a sufficiently
long time all of the particles will reach the ground state so that
the system energy is minimal. The simulated annealing bears an
analogy to the physical process. As the initial condition, the
control parameter (analogy to the temperature) adopts a proper
value. Then one needs to carefully tune the parameter according to
certain cooling schedule. At each parameter value, by devising an
appropriate neighborhood generation (or state updating) mechanism
and acceptance criterion, the transitions between the accepted
states prove to form a homogeneous Markov chain. And the global
optimal state(s), which minimize(s) or maximize(s) the cost
function (analog to the system energy), will be achieved as the
control parameter tends to zero. For detail, see, for example,
\cite{Aarts simulated,Laarhoven simulated}.

As we have mentioned previously, surrogates produce by {\it
Algorithm 2} preserve the amplitude distribution of the original
data but differentiate in the power spectra. One remedy for this
problem is to iterate the surrogate generation procedures for a
number of times until certain criterion is satisfied. However,
usually there is no guarantee that the chosen criterion is the
best, therefore it is possible that the iterative algorithm only
engenders sub-optimal solutions (i.e., local minima) or even
worse. For this reason, applying the simulated annealing method
for surrogate generation, in contrast, will often achieve better
performance, as pointed out by Schreiber \cite{Schreiber
constrained}. However, in some situations, this approach can be
time consuming and provide limited benefit \cite{Small applied}.

Because the linear autocorrelations are directly related to the
power spectrum \cite{Box time}, in configuration of the simulated
annealing it is natural to choose, as the cost function, the norm
of the difference between the linear autocorrelations of the
original and the simulated data (see Eq.(2) of \cite{Schreiber
constrained}). The (accepted) simulated data is updated by
exchanging the pairs of the former one, while the expected
surrogate is the final simulated data when the stopping criterion
is reached \footnote{To save time, we skip the introduction of the
configuration of other components like the initialization of the
control parameter, the acceptance criterion, the cooling schedule
and the stopping criterion, the readers are referred to the work
\cite{Schreiber constrained} and the references therein for more
detail.}. Hence it is easy to arrive the conclusion that thus
generated surrogates are also constrained realizations in the
sense that they preserve the amplitude distribution of the
original data.

In practice, one inherent advantage to apply the simulated
annealing method is that, after adequate cooling for the control
parameter, the obtained solution could reach a local minimum
sufficiently close to the global one. Another advantage is that it
does not need to invert the nonlinear filter for the test of {\it
NH2}. Of course, there is also one obvious disadvantage, that is,
depending on the size of the problem and the configuration of the
algorithm, the computational time might substantially increase as
often the case.

\section{Surrogate test for detection of aperiodicity}

In the previous section we have described the tests that can be
applied to arbitrary time series data. We now confine our
discussion to pseudo-periodic data. By {\it pseudo-periodic data}
we mean those time series that exhibit strong periodic trends
manifesting as clear spikes in the frequency domain \cite{Small
applied} (see Figure \ref{somedata}(a) and (c) ). The underlying
systems of pseudo-periodic data can be periodic orbits
contaminated with observational or dynamical data, or oscillatory
chaotic systems (for example the R\"{o}ssler system). In this
sense, one could also apply the surrogate data method to detect
chaos in a pseudo-periodic time series, which will be the focus of
this section.

\subsection{The cycle shuffled algorithm}

Here let us first specify the null hypothesis, which assumes that
there are no temporal correlations at all between the
spike-and-wave patterns (i.e., the individual cycle patterns) of
the pseudo-periodic time series \cite{Theiler on}. Obviously, any
purely periodic time series is consistent with the hypothesis.
However, if there exists perturbations to the periodic orbits,
then it requires that those inter-cycle perturbations are also
uncorrelated at all (for example, the i.i.d noise), which is a
stronger constraint than that of the null hypothesis to be
introduced in the next subsection.

Since there are no temporal correlations between individual cycle
patterns, similar to the idea of {\it block bootstrap}
\cite{Kunsch jackknife} to decompose and shuffle individual
blocks, the natural way to produce pseudo-periodic surrogates is
to first extract the individual cycles from the pseudo-periodic
time series and then randomly shuffle these cycles. For this
reason, this method is often called cycle shuffled algorithm. Note
that, although not explicitly specified in the null hypothesis, in
order to let the algorithm produce constrained realizations, it
requires that intra-cycle dynamics of the individual cycles
distributes periodically.

Theoretically the cycle shuffled algorithm is very simple, but
there is a practical problem in implementation, which essentially
lies in the difficulty in extracting the individual cycles from
the test data. Given a pseudo-periodic time series, shuffling the
split cycles will often lead to the spurious discontinuity. To
eliminate this phenomenon, one could vertically shift the
individual cycles, but it often turns out to make the data become
non-stationary and thus generate artificial long term correlation
(see, for example, the illustrations in \cite{Small applied}). In
the following we will introduce another algorithm that produces
surrogates from a different viewpoint and avoids the above
problems. Moreover, if one is interested primarily in the
variation between cycles, it may be better to study that directly
\cite{ZhangJei}.

\subsection{The temporal shift algorithm}

Here the null hypothesis under test is that the pseudo-periodic time
series is produced from a periodic orbit perturbed by noise
components that are identically distributed and uncorrelated for
large enough temporal shifts \cite{Luo surrogate}. This null
hypothesis is slightly more general than that in the previous
subsection in the sense that, it does not require that there are
no temporal correlations between the individual cycles
\footnote{As an example, one may consider the case of a periodic
orbit contaminated by linear colored noise, which is consistent
with the hypothesis presented here but not the former one if the
characteristic decorrelation time of the color noise is larger
than the length of the data period.}.

\begin{figure}[!t]

\centering

\includegraphics[width=4.5in]{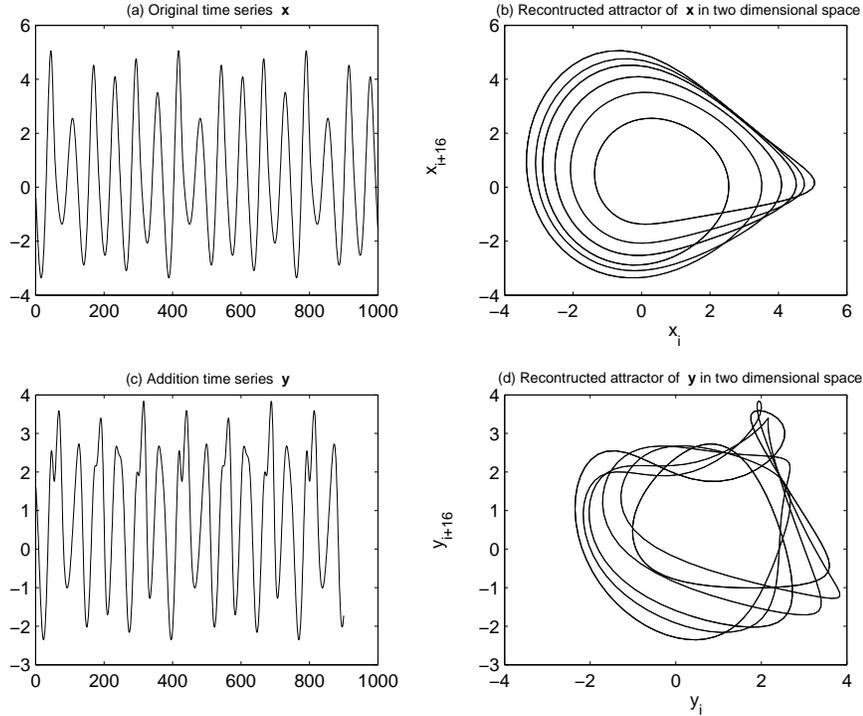}

\caption{An illustration of that the addition of two subsets of a
periodic time series does not preserve the original cycle pattern.
Panel (a) Original period 6 time series ${\bf
x}=\{x_i\}_{i=1}^{1000}$ from the R\"ossler system; (b) The
reconstructed attractor in two dimensional embedding space
$x_{i+16}$ vs. $x_i$; (c) Addition time series ${\bf
y}=\{y_i:y_i=x_i+x_{i+100}\}_{i=1}^{900}$; (d) The reconstructed
attractor in two dimensional embedding space $y_{i+16}$ vs. $y_i$.
From the reconstructed attractor \cite{Mane on,Sauer embed,Takens
strange} in panel (d), one can find that the addition time series
is also period 6, however its cycle pattern differs from that of
the original time series.}

\label{PeriodTS}

\end{figure}

Note that, adding together two subsets of the same periodic time
series will lead to a new periodic data, while for chaotic time
series applying the same transformation will usually increase the
complexity. With this property, one could adopt the temporal shift
algorithm as well to produce the surrogates under the hypothesis,
i.e., given a $N$-point pseudo-periodic time series ${\bf
d}=\{d_i\}_{i=1}^N$, one could generate the surrogates ${\bf s}$
according to the formula $s_i=\alpha d_i+\beta d_{i+\tau}
(i=1,2,...,N-\tau)$ with proper coefficient ratios $\alpha/\beta$.
However, there is still an important difference, that is, for
pseudo-periodic time series usually the temporal shift algorithm
is not constrained. This is because in this situation the
surrogates only preserve the periodicity but not necessarily the
cycle pattern (see Fig. \ref{PeriodTS} for an illustration).
Therefore the surrogates, although also periodic, may not come
from the same underlying system as that of the original data. With
this consideration, one shall choose a pivotal test statistic with
robust performance against noise in calculation. An example of
such choices is the correlation dimension evaluated by the
Gaussian kernel algorithm (GKA), as described in \cite{Diks
estimating,Yu efficient}.

\subsection{The attractor trajectory surrogate algorithm}

The attractor trajectory surrogate (ATS) algorithm produces
surrogates by inferring the underlying systems from a local model,
and contaminating a trajectory on the attractor with dynamical
noise \cite{Small applied}. In this way, the surrogates preserve
the gross scale dynamics of the original data but destroy the fine
scale one.

Examples of the ATS algorithm can be found in \cite{Dolan
surrogate,Small surrogate,Unsworth a}. Here we only introduce the
pseudo-periodic surrogate (PPS) algorithm \cite{Small surrogate}
which is designed for detection of the null hypothesis that
assumes the time series from a periodic orbit with uncorrelated
dynamical noise.

Given a scalar time series ${\bf z}=\{z_i\}_{i=1}^N$, the procures
for surrogate generation go as follows \cite{Small surrogate,Small
applied}:

\begin{enumerate}

\item Choose proper embedding dimension $d_e$ and time delay
$\tau$ for time delay embedding reconstruction \cite{Mane on,Sauer
embed,Takens strange}. By reconstruction based on the original
data ${\bf z}$, one obtains a set of delay vectors ${\bf V}=\{{\bf
v}_i\}_{i=1}^{d_w}$ with delay vector ${\bf
v}_i=[z_i,z_{i+1},...,z_{i+(d_e-1)\tau}]^T$ and the embedding
window $d_w=N-(d_e-1)\tau$.

\item Randomly choose a delay vector ${\bf \chi}_0 \in {\bf V}$
for initialization.

\item Let index $k$ start from $k=1$.

\item Let ${\bf \chi}_k$ be the current delay vector in operation.
Search in ${\bf V}$ the neighbors of ${\bf \chi}_k$ and randomly
pick out one as the successor of ${\bf \chi}_k$, which is denoted
by ${\bf \chi}_{k+1}$.

\item Take ${\bf \chi}_{k+1}$ as the current operation vector.
Repeat the procedure in step 4 until index $k$ reaches the
specified length, say, $M$.

\item The surrogate data ${\bf s}=\{s_i:s_i=(\chi_i)_1\}_{i=0}^M$,
where $(\chi_i)_1$ denotes the first element in vector $\chi_i$.

\end{enumerate}

It was shown \cite[p.160]{Small applied} that the surrogates ${\bf
s}$ produced through the above procedures share the same vector
field as the original data ${\bf z}$ but are contaminated with
dynamical noise. However, the produced surrogates may not strictly
preserve the gross scale dynamics of the original data as we
observed in practice. In this sense, the surrogates are not
constrained realizations, therefore in tests one needs to choose a
pivotal statistic (e.g., the correlation dimension as
aforementioned).

\section{Surrogate test for detection of correlations between irregular fluctuations}

A new application of the surrogate data method is to detect
correlations between irregular fluctuations possibly with a long
term trend \cite{Nakamura shuffle}. The corresponding null
hypothesis is that irregular fluctuations are independently
distributed, which differentiates {\it NH0} of Section 2 in that
it does not require the identical distribution of the
fluctuations.

Similar to the idea of the attractor trajectory surrogate (ATS)
algorithm, the surrogate generation algorithm devised in
\cite{Nakamura shuffle} also aims to preserve the global behavior
(e.g., the trend) but destroy the local one. In the following let
us explain in more detail.

Given a scalar data ${\bf d}=\{d_i\}_{i=1}^N$, let the index set
be ${\bf h}=\{h_i:h_i=i\}_{i=1}^N$ such that $d_i=d_{h_i}$, then
the concrete steps for implementation of the idea include:

\begin{enumerate}

\item Perturb the original index set ${\bf h}$ with Gaussian
random numbers $\{\xi_i\}_{i=1}^N$ so as to obtain a real number
set ${\bf r}=\{r_i: r_i=h_i+A \xi_i\}_{i=1}^N$.

\item Sort ${\bf r}$ in the ascendant order to produce a new data
set ${\bf t}$. Re-ordering the index set ${\bf h}$ will lead to
the disturbed index set ${\bf k}=\{k_i\}_{i=1}^N$, which satisfies
$t_i=r_{k_i}$.

\item The surrogate data ${\bf s}=\{s_i\}_{i=1}^N$ is obtained by
letting $s_i=d_{k_i}$.

\end{enumerate}

By choosing a proper amplitude $A$, typically the irregular
fluctuations will only slightly move from the positions in the
original data, therefore the generation mechanism is called
small-shuffle surrogate (SSS) algorithm. But note that, although
the surrogates ${\bf s}$ preserve the amplitude distribution of
the original data ${\bf d}$, usually the surrogates are not
constrained realizations. This is because the irregular
perturbations are possibly not identically distributed, and
locally shuffling the irregular perturbations may not exactly
preserve the global dynamics. Thus, one needs to carefully choose
the test statistics, which may be the linear autocorrelation
function or the average mutual information as suggested in
\cite{Nakamura shuffle}.

\section{Summary}

In the above sections we have reviewed the concept of surrogate
data tests, the primary components that form the framework of this
method, and some important properties of these components. We have
also reviewed the applications of the surrogate data method, with
the emphasis on some recently developed surrogate generation
algorithms.

In all of the applications, since the specified null hypotheses
are composite, it is required that the surrogate generation
algorithms are nonparametric and work for any process consistent
with the hypothesis. Consequently, the broader range of the
underlying processes a composite null hypothesis may cover, the
more difficult it is to design the corresponding nonparametric
surrogate algorithm. This fact limits the applications of the
surrogate data method to detect many other interesting properties.

Another challenge is the design of a proper test statistic. The
problem comes not only from the requirement of pivotal-ness when
the surrogates are not constrained realizations, but also from the
expectation that one obtains the exact confidence level to reject
a null hypothesis. Although simple in handle, the two
discriminating criteria described in Section 2 actually cannot
lead to inferences with exact confidence levels (see the
discussion in \cite{Luo nonparametric}). The solution to this
problem requires that one seeks the full knowledge of the
distribution of the test statistic, which is often infeasible,
even only at the asymptotic level, for many nonlinear statistics.

\section{Acknowledgements*}

This research was supported by Hong Kong University Grants Council
Competitive Earmarked Research Grant~(CERG) No. PolyU 5235/03E.

\end{document}